\documentclass[epj,final]{svjour}
\usepackage{graphics}
\begin{document}
\title{Fine and hyperfine interaction on the light cone}
\author{Hans-Christian Pauli}
\institute{Max-Planck-Institut f\"ur Kernphysik, D-69029 Heidelberg }
\date{20 December 2003}  
\abstract{%
   The formalism for the spin interactions in the front form (light-cone)
   is re-phrased in terms of an instant form formalism.
   It is shown how to unitarily transform the Brodsky-Lepage spinors
   to Bj\o rken-Drell spinors and to re-phrase the so called
   spinor matrix in terms of the interactions one is familiar with
   from atomic and Dirac theory.~---
   One retrieves the (relativistic) kinetic correction, 
   the hyperfine and the Darwin term which acts even when 
   wave function is spherically symmetric. One also retrieves
   angular momentum dependent terms like the spin-orbit interaction
   in a relativistically correct way; and one obtains additional
   terms which thus far have not been reported particularly various
   $\vec L^2$-dependent terms. Since the approach includes
   the full retardation, one gets additional, thus far unknown terms.
   The differ from atomic and Dirac theory, since there only
   that part of the vector potential is usually included 
   which is generated by the atomic nucleus.
   Quite on purpose, the paper is kept formal.~---
} 
\maketitle
\section{The light-cone integral equation}
This paper number 3 in a row of 3 \cite{Pau03a,Pau03b} on the bound state
problem in gauge theory \cite{BroPauPin98} deals with the
technical question of how to formulate the fine and hyperfine interaction
in the one-body integral `master' equation 
which has been previously derived \cite{Pau03b,BroPauPin98}.

I therefore jump immediately to Eq.(16) of \cite{Pau03b},
\begin{eqnarray} 
\lefteqn{\hspace{-2em}
    M^2\psi_{h_1h_2}(x,\vec k_{\!\perp}) = 
    \left[ 
    \frac{m^2_{1} + \vec k_{\!\perp}^{\,2}}{x} +
    \frac{m^2_{2} + \vec k_{\!\perp}^{\,2}}{1-x}  
    \right]
    \psi_{h_1h_2}(x,\vec k_{\!\perp})  
}\nonumber\\ 
    &-& {1\over 4\pi^2}
    \sum _{ h_1',h_2'} \!\int\!\displaystyle 
    \frac{dx^\prime d^2 \vec k_{\!\perp}^\prime
    \;\psi_{h_1'h_2'}(x',\vec k_{\!\perp}')}
    {\sqrt{ x(1-x) x'(1-x')}}
    \frac{\alpha_c(Q)}{Q^2} R(Q) 
\nonumber\\  
    &\times&
    \left[\overline u(k_1,h_1)\gamma^\mu u(k_1',h_1')\right]
    \left[\overline v(k_2',h_2')\gamma_\mu v(k_2,h_2)\right] 
\,.\label{eq:1}\end {eqnarray}
Here, $M ^2$ is the eigenvalue of the invariant-mass squared. 
The associated eigenfunction $\psi_{h_1h_2}(x,\vec k_{\!\perp})$ 
is the probability amplitude 
$\langle x,\vec k_{\!\perp},h_{1};1-x,-\vec k_{\!\perp},h_{2}
\vert\Psi_{q\bar q}\rangle$ 
for finding the quark with momentum fraction $x$, 
transversal momentum $\vec k_{\!\perp}$ and helicity $h_{1}$,
and correspondingly the anti-quark.
Their (effective) masses are denoted by $m _1$  and $m _2$,
and $u(k_1,h_1)$ and $v(k_2,h_2)$ are their 
Dirac spinors in Lepage Brodsky convention,  
as given in \cite{BroPauPin98}. 
The (effective) coupling function 
$\alpha_c(Q)=\frac43\overline\alpha(Q)$ is also given 
in \cite{BroPauPin98}. The kernel is governed by the 
mean four-momentum transfer,
$Q^2=\frac12\left(Q_{q}^2+Q_{\bar q}^2\right)$,
where  
\begin{eqnarray}
   Q_{q}^2 = -(k_1-k_1')^2
\quad\mbox{ and }\quad
   Q_{\bar q}^2 = -(k_2-k_2')^2
\,,\label{neq:2}\end{eqnarray}
are the Feynman four-momentum transfers of quark and anti-quark, 
respectively.
The regulator function $R(Q)$, finally,  
removes the ultraviolet singularities and regulates the interaction.
Note that the equation is fully relativistic and covariant.
It coincides literally with Eq.(4.101) of \cite{BroPauPin98}.
   
Krautg\"artner \textit{et al} \cite{KraPauWoe92}
and Trittmann \textit{et al} \cite{TriPau00} have shown 
how to solve such an equation numerically with high precision. 
But since the numerical effort is considerable, 
it is reasonable to work first with simpler models.
The aim of the present work is to derive such ones. 

The aspects of regularization and renormalization 
have been emphasized in \cite{Pau03a,Pau03b}, 
resulting in an explicit construction of the regulator function $R(Q)$.
The case was worked out within the so called Singlet-Triplet model. 
Here, I address to go beyond that, particularly to derive
a model for the spin-orbit interaction, 
which had been suppressed on purpose in \cite{Pau03b}. 

\section{Transforming the integral equation}
The light-cone integral equation (\ref{eq:1}) has the unpleasant
aspect that the integration variables have a completely
different support,
\begin{eqnarray*}
   0 < &x& < 1
\,,\qquad 
   -\infty < \vec k_{\!\perp} < +\infty 
\,.\end{eqnarray*}
Therefore, practically in all of the numerical work particularly
in \cite{KraPauWoe92} and \cite{TriPau00}, the variable transform 
\begin{eqnarray}
   x(k_z)  &=& \frac{E_1+k_z}{E_1+E_2}
\,,\label{eq:2}\\ \mbox{ with }
   E_{1,2} = E_{1,2} (k) &\equiv&
   \sqrt{m^{\,2}_{1,2}+ k_z^2 + \vec k_{\!\perp}^{\,2}} 
\,,\nonumber\end{eqnarray}
has been used to transform to integration variables 
\begin{eqnarray*}
   -\infty < &k_z& < +\infty
\,,\qquad 
   -\infty < \vec k_{\!\perp} < +\infty 
\,,\end{eqnarray*}
with the same support. 
While $k_z$ varies from $-\infty$ to $+\infty$, the $x(k_z)$  
varies from $0$ to $1$. 
The particles are then described by their front form four-momenta 
\begin{eqnarray*}
  \begin{array} {lccl c lccl }
     k^+_1 &=& &k_z + E_1(k)\,,&\quad& k^+_2 &=&-&k_z + E_2(k)\,, \\
     {\vec k_{\!\perp}}_1 &=& &  \vec k_{\!\perp}\,,&\quad&
     {\vec k_{\!\perp}}_2 &=&-&  \vec k_{\!\perp}\,,   \\
     k^-_1 &=&-&k_z + E_1(k)\,,&\quad& k^-_2 &=& &k_z + E_2(k)\,. \\
  \end{array}
\end{eqnarray*}
Or, they are described by the instant form four-momenta 
\begin{eqnarray*}
  \begin{array} {lccl c lccl }
     k^0_1 &=& &E_1(k)\,,&\quad& k^0_2 &=& & E_2(k)\,, \\
     {\vec k}_1 &=& &  \vec k\,,&\quad&
     {\vec k}_2 &=&-&  \vec k\,,   
  \end{array}
\end{eqnarray*}
with $\vec k \equiv(\vec k_{\!\perp},k_z)$. Such a switching between
front form and instant form parameterization is possible, since the
four-vectors of the constituents refer to free particles. 
The free invariant mass of the two particles,  
\begin{eqnarray}
   M^2_\mathrm{free} =
   \frac{ m^2_{1} + \vec k_{\!\perp}^{\,2}}{x} +
   \frac{ m^2_{2} + \vec k_{\!\perp}^{\,2}}{1-x} = 
   \left(E _1(k) + E _2(k)\right)^2
\,,\end{eqnarray}
looks like in the rest frame of the instant form ($\vec P=0$).
For vanishing $k$ it is $\left(m_1+ m_2\right)^2$.
The discrepancy can be calculated exactly as \cite{Pau00c}
\begin{eqnarray*}
\lefteqn{\hspace{-2em}
   \left(E _1 + E _2\right)^2 - \left(m_1+ m_2\right)^2=   
}\nonumber\\ & &
   \left(E_1 + E_2 - m_1 - m_2\right)
   \left(E_1 + E_2 + m_1 + m_2\right)
\,,\end{eqnarray*}
and therefore as
\begin{eqnarray*}
\lefteqn{\hspace{-2em}
   \left(E _1 + E _2\right)^2 - \left(m_1+ m_2\right)^2=   
}\nonumber\\ & &
   \left(\frac{E_1^2-m_1^2}{E_1+m_1} +
   \frac{E_2^2-m_2^2}{E_2+m_2} \right)
   \left(E_1 + E_2 + m_1 + m_2\right)
\,,\end{eqnarray*}
With the reduced mass,
\begin{eqnarray}
   \frac{1}{m_r} = \frac{1}{m_1} + \frac{1}{m_2}
\,,\qquad
   m_r= \frac{m_1 m_2}{m_1 + m_2}
\,,\label{eq:5}\end{eqnarray}
and the dimensionless $A(k)$ and $B(k)$,
\begin{eqnarray} \textstyle 
    A(k) &\equiv&   \textstyle 
    m_r \frac{E_1(k)+E_2(k)} {E_1(k)E_2(k)}
\,,\nonumber\\  \textstyle
   B(k) &\equiv&  \textstyle 
   \frac{E_1(k) + m_1 + E_2(k) + m_2}
   {\left(m_1 + m_2\right)}
\nonumber\\  \textstyle &\times& \textstyle 
   \Big(\frac{m_r}{E_1(k)+m_1}+\frac{m_r}{E_2(k)+m_2}\Big)
\,,\label{neq:5}\end{eqnarray}
the free invariant mass and the exact Jacobian of the transformation 
(\ref{eq:2}) is therefore
\begin{eqnarray}  
    M^2_\mathrm{free} &=& 
    (m_1+ m_2)^2 + (m_1+ m_2) \frac{k ^2}{m_r}\ B(k)
\,,\label{neq:6}\\
    dx &=&  x(1-x)\frac {dk_z}{m_r}\ \frac {1}{A (k)} 
\,,\end{eqnarray}
respectively, see also \cite{Pau00c}.

The transformation from $(x,\vec k_{\!\perp})$ to $(k_z,\vec k_{\!\perp})$
will be presented in two steps. 
In the first step, the variables are transformed and 
Eq.(\ref{eq:1}) becomes 
\begin{eqnarray*} 
\lefteqn{\hspace{-2em}
   \left[M^2-\left(E _1(k) + E _2(k)\right)^2\right]
   \psi_{h_1h_2}(k_z,\vec k_{\!\perp}) = 
}\nonumber\\ 
    &-& \sum _{ h_1',h_2'} \!\int\!
    dk'_z d^2 \vec k'_{\!\perp}
    \frac{\sqrt{x'(1-x')}}{\sqrt{x(1-x)}}
    \frac{\psi_{h_1'h_2'}(k_z',\vec k_{\!\perp}')}{A (k')}
\nonumber\\  
    &\times&
    \frac{\alpha_c(Q)}{Q^2} \frac{R(Q)}{4\pi^2 m_r} 
\nonumber\\  
    &\times&
    \left[\overline u(k_1,h_1)\gamma^\mu u(k_1',h_1')\right]
    \left[\overline v(k_2',h_2')\gamma_\mu v(k_2,h_2)\right] 
\,.\end {eqnarray*}
One notes that the kernel is not symmetric under the exchange of 
primed ($'$) and unprimed quantities, as opposed to Eq.(\ref{eq:1})
which is symmetric. This asymmetry can however be removed as usual, 
by multiplying the equation in a second step
with $\sqrt{x(1-x)}/\sqrt{A(k)}$, 
\begin{eqnarray} 
\lefteqn{\hspace{-2em}
    \left[M^2-\left(E _1 + E _2\right)^2\right]
    \overbrace{
    \frac{\sqrt{x(1-x)}}{\sqrt{A(k)}}
    \psi_{h_1h_2}(k_z,\vec k_{\!\perp})
    }^{\phi_{h_1h_2}(k_z,\vec k_{\!\perp})} = 
}\nonumber\\ 
    &-& \sum _{ h_1',h_2'} \!\int\! dk'_z d^2 \vec k'_{\!\perp}
    \frac{\sqrt{x'(1-x'}}{\sqrt{A(k')}}
    \psi_{h_1'h_2'}(k_z',\vec k_{\!\perp}')
\nonumber\\  
    &\times&
    \frac{\alpha_c(Q)}{Q^2} \frac{R(Q)}{4\pi^2 m_r} 
    \frac{1}{\sqrt{A(k)A(k')}}
\nonumber\\  
    &\times&
    \left[\overline u(k_1,h_1)\gamma^\mu u(k_1',h_1')\right]
    \left[\overline v(k_2',h_2')\gamma_\mu v(k_2,h_2)\right] 
\,.\label{eq:6}\end {eqnarray}
In the numerical work \cite{KraPauWoe92,TriPau00}, 
the reduced wave function $\phi_{h_1 h_2}(k_z,\vec k_{\!\perp})$
is calculated first. It is then converted 
to $\psi_{h_1 h_2}(x,\vec k_{\!\perp})$ by  
\begin{eqnarray} 
   \psi_{h_1 h_2}(x,\vec k_{\!\perp})= 
   \frac{\sqrt{A(k_z(x),\vec k_{\!\perp})}}{\sqrt{x(1-x)}}
   \phi_{h_1 h_2}(k_z(x),\vec k_{\!\perp})
\,,\label{eq:7}\end{eqnarray}   
\textit{i.e.} by the substitution $k_z=k_z(x)$ 
which is inverse to Eq.(\ref{eq:2}).
Krautg\"artner \cite{KraPauWoe92}
and particularly Trittmann \cite{TriPau00} have presented 
beautiful three-dimensional plots of $\psi_{h_1 h_2}$.

The variable transformation (\ref{eq:2}) 
is applied here in a strict mathematical sense. 
It does not change the physical content, particularly
not the eigenvalue spectrum $M^2_i$.
The transformed integral equation (\ref{eq:6}) 
looks like an equation in usual momentum space.
But one should emphasize that it continues to be a front form equation 
with the sole purpose to generate $\psi_{h_1 h_2}(x,\vec k_{\!\perp})$.
The reduced wave function $\phi_{h_1 h_2}(x,\vec k_{\!\perp})$
has no physical interpretation.

The kernel of Eq.(\ref{eq:6}) depends on $|\vec k'|^2$ and 
Lorenz-invariants like $Q^2$ or $[\gamma^\mu][\gamma_\mu]$.
It is therefore invariant under spatial rotations.
The occurrence of spin-degenerate multiplets in the numerical
solutions of \cite{TriPau00} become thus understandable 
\textit{e posteriori}. 
It seems as if all light-cone specific troubles with
rotations of the coordinate system \cite{BroPauPin98} are absorbed
in the $x(1-x)$-factor in Eq.(\ref{eq:7}):
The reduced wave function 
$\phi_{h_1 h_2}(k_z,\vec k_{\!\perp})$ transforms covariantly under rotations
while $\psi_{h_1 h_2}(x,\vec k_{\!\perp})$ does not.

\section{The Melosh rotated integral equation}
Much of the difficulty in getting the reduced wave function 
in practice \cite{KraPauWoe92,TriPau00}, 
and to understand the structure of the numerical solutions, 
resides in the so called spinor factor 
\begin{eqnarray} 
\lefteqn{\hspace{-0em}
   \langle h_1,h_2|S|h'_1,h'_2\rangle = 
}\nonumber\\ &=&
   \left[\overline u(k_1,h_1)\gamma^\mu u(k_1',h_1')\right] 
   \left[\overline v(k_2',h_2')\gamma_\mu v(k_2,h_2)\right]  
\,.\end{eqnarray}
The $4\times4$ matrix in the helicities can be 
simplified by 
\begin{eqnarray}
   \left[\overline v(k_2',h_2')\gamma_\mu v(k_2,h_2)\right] =
   \left[\overline u(k_2,h_2)\gamma_\mu u(k_2',h_2')\right]
\,.\end{eqnarray}
This general spinor identity allows to work only the $u$-spinors.
But even then, the spinor factor
\begin{eqnarray} 
\lefteqn{\hspace{-2em}
   \langle h_1,h_2|S|h'_1,h'_2\rangle \equiv 
   \langle h_1,h_2|S(x,\vec k_{\!\perp};x',\vec k'_{\!\perp})|h'_1,h'_2\rangle = 
}\nonumber\\ & &
   \left[\overline u(k_1,h_1)\gamma^\mu u(k_1',h_1')\right] 
   \left[\overline u(k_2,h_2)\gamma_\mu u(k_2',h_2')\right]
\,,\label{eq:10}\end{eqnarray}
is a terribly complicated function of helicities and light-cone
momenta $x,\vec k_{\!\perp}$ and $x',\vec k'_{\!\perp}$,
as seen in the explicit tables in \cite{Pau00c}.
They become even more complicated if 
one expresses them in terms of $k_z,\vec k_{\!\perp}$ and 
$k'_z,\vec k'_{\!\perp}$, see \cite{TriPau00}. 

One conjectures that the spinor function (\ref{eq:10}) 
is much simpler if the Lepage-Brodsky spinors 
$u(k,h)\equiv u^{\scriptstyle LB}(k,h)$ 
are replaced by the 
Bj\o rken--Drell spinors $u^{\scriptstyle BD}(k,s)$ \cite{BjD64}.
As to be seen, this is the case indeed.

The way this can be done was shown first by 
Krassnigg \textit{et al.} \cite{KrassPau02}, 
and their work shall be repeated here in short.~-- 
Both, the Lepage-Brodsky and the Bj\o rken--Drell spinors 
are solutions to the same equation, the free Dirac equation
$(\not p-m)\, u(p,\lambda)=0$.
Hence, they must be linear superpositions of each other. We define
\begin{equation}
   u ^{LB} _\alpha (k,h) = \sum _s u ^{BD} _\alpha (k,s)   
   \langle s|\omega|h\rangle  
\,.\label{eq:11}\end{equation}
The transformation matrix $\langle s|\omega|h\rangle$ 
is independent of the Dirac indices $\alpha$.
If both spinors have the same normalization,
the transformation is unitary~--- and then called a Melosh rotation 
\cite{Mel74,Dzi87,AhS92}.

The Lepage-Brodsky spinors are conventionally  
normalized as $\overline u ^{LB}(k,h) u^{LB} (k,h') = 2m \delta_{hh'}$,
while the Bj\o r\-ken--Drell spinors are normalized to unity. 
We change convention of the Lepage-Brodsky spinors 
by requiring
\begin{equation}
   \overline u^{LB}(k,h) u^{LB}(k,h') = \delta_{hh'}
\,.\end{equation}
Eventually, this implies to multiply the 
kernel of Eq.(\ref{eq:6}) with the factor $4m_1m_2$. 

The expansion coefficients from Eq.(\ref{eq:11}) are then determined by
\begin{eqnarray}
   \langle s|\omega|h\rangle  = 
   \sum_{\alpha}
   \overline u^{BD}_\alpha(k,s)\ u^{LB}_\alpha(k,h) 
\,.\label{eq:13}\end{eqnarray}
The Lepage-Brodsky spinors are \cite{BroPauPin98}:
\begin{eqnarray*}
   u^{\scriptstyle LB}(k,h) = 
   \frac{1}{\sqrt{4m k^+}}
   \left(\begin{array}{l|l}
   \quad h:\uparrow & \quad h:\downarrow \\ \hline 
   k^+ + m &  -k_l \\ k_r &  \phantom{-}k^+ + m \\ k^+ -m &  \phantom{-}k_l \\ 
   k_r & -k^+ +m  
   \end{array}\right) 
\,,\end{eqnarray*}
with $k_r \equiv k_x+ik_y$ and $k_l \equiv k_x-ik_y$.
The Bj\o rken--Drell spinors are \cite{BjD64}:
\begin{eqnarray*}
   u^{\scriptstyle BD}(k,s)  = 
   \frac{1}{\sqrt{2m(E + m)}} 
   \left(\begin{array}{l|l}
   \quad s:\uparrow & \quad s:\downarrow \\ \hline
   E + m & \phantom{-}0  \\ 0   & \phantom{-}E + m \\ 
   k_z   & \phantom{-}k_l\\ k_r & -k_z 
   \end{array}\right) 
\,,\end{eqnarray*}
with $E\equiv E(k)$ as in Eq.(\ref{eq:2}).
The four overlap matrix elements of Eq.(\ref{eq:13}) are then 
calculated as
\begin{equation}
   \langle s|\omega|h\rangle  = 
   \frac{1}{\sqrt{2p^+(E+m)}}\left(
   \begin{array}{@{}ll@{}} 
     k^+ + m &           -k_l    \\ 
     k_r     & \phantom{-}k^+ + m 
   \end{array}\right) 
\,,\end{equation}
with the rows labeled by $s$ and the columns by $h$.
One calculates
\begin{eqnarray*}
\lefteqn{\hspace{-1em}
   \sum _{h}\langle s|\omega|h\rangle \langle h|\omega^\dagger|s'\rangle  = 
   \frac{1}{2k^+(E+m)} 
}\nonumber\\ &\times& \left(
   \begin{array}{@{}ll@{}}
    k^+ + m & -k_l    \\ 
    k_r     &  k^+ + m 
   \end{array}
   \right) \left(
   \begin{array}{@{}ll@{}}  
    k^+ + m & k_l    \\ 
   -k_r     & k^+ + m 
   \end{array}\right) =
   \left(\begin{array}{@{}ll@{}} 1 & 0 \\ 0 & 1 \end{array}\right) 
\,,\end{eqnarray*}
since $(p^+ +m)^2+ p_l p_r = 2p^+(E+m)$, verifying this way that
the transformation is unitary, indeed. 

Since the spinors appear in bilinear combinations,
it is convenient to define the unitary direct product 
\begin{equation} 
   \langle s_1 s_2|\Omega|h_1 h_2\rangle =
   \langle s_1|\omega|h_1\rangle \otimes
   \langle s_2|\omega|h_2\rangle
\,.\end{equation}
Introducing a second reduced wave function $\varphi_{s_1s_2}$ by
\begin{eqnarray} 
   \varphi_{s_1s_2}(\vec k) = \sum _{ h_1,h_2} 
   \langle s_1 s_2|\Omega|h_1 h_2\rangle  \phi_{h_1h_2}(\vec k) 
\,,\end {eqnarray}
Eq.(\ref{eq:6}) can be transformed unitarily to
\begin{eqnarray} 
\lefteqn{\hspace{-0em}
    \left[M^2-\left(E _1(k)+E _2(k)\right)^2\right]
    \varphi_{s_1s_2}(\vec k) = 
}\nonumber\\ 
    &-& \frac{(m_1+m_2)}{\pi^2}
    \sum _{ s_1',s_2'} \!\int\! dk'_z d^2 \vec k'_{\!\perp}
    \frac{\varphi_{s_1's_2'}(\vec k')}{\sqrt{A(k)A(k')}}
    \frac{\alpha_c(Q)}{Q^2} R(Q) 
\nonumber\\  
    &\times&
    \left[\overline u(k_1,s_1)\gamma^\mu u(k_1',s_1')\right]
    \left[\overline u(k_2,s_2)\gamma_\mu u(k_2',s_2')\right] 
\,.\label{eq:19}\end {eqnarray}
Once one has the wave functions 
$\varphi_{s_1s_2}(k_z,\vec k_{\!\perp})$,
one can unitarily transform them back to the light-cone wave functions by
\begin{eqnarray} 
\lefteqn{\hspace{-2em}
   \psi_{h_1 h_2}(x,\vec k_{\!\perp})= 
   \frac{\sqrt{A(k_z(x),\vec k_{\!\perp})}}{\sqrt{x(1-x)}} 
}\label{eq:21}\\ &\times&
   \sum_{s_1,s_2}
   \langle h_1 h_2|\Omega^\dagger(k_z(x),\vec k_{\!\perp})|s_1 s_2\rangle 
   \varphi_{s_1 s_2}(k_z(x),\vec k_{\!\perp})
\,,\nonumber\end{eqnarray}   
in analogy to Eq.(\ref{eq:7}).

The spinors in Eq.(\ref{eq:19}) are Bj\o rken--Drell spinors.
On the technical level, they are much more transparent
than those of Lepage--Brodsky, as to be seen next. 
The previous numerical work \cite{KraPauWoe92,TriPau00} done 
with Eq.(\ref{eq:19}) would have been much easier than 
with Eq.(\ref{eq:6}). 
But at that time, the present physical insight was lacking.

\subsection{Definition of the spinor factor}
The spin dependence in Eq.(\ref{eq:19}) resides in the
spinor factor  
\begin{eqnarray*} 
\lefteqn{\hspace{-0em}
   \langle s_1,s_2|S|s'_1,s'_2\rangle = 
   \left[\overline u(k_1,s_1)\gamma^\mu u(k_1',s_1')\right] 
}\nonumber\\ && \hspace{15ex}\times
   \left[\overline u(k_2,s_2)\gamma_\mu u(k_2',s_2')\right]
\,.\end{eqnarray*} 
With
$\gamma^\mu(1)\gamma_\mu(2)= \gamma^0(1)\gamma^0(2)-\gamma^i(1)\gamma^i(2)$,  
where  
\begin{eqnarray*} 
   \gamma^0 &=&
   \left(\begin{array}{@{}rr@{}} 1&0\\ 0&-1 \end{array}\right)
   \,,\qquad
   \{\gamma^i\}=\vec\gamma=
   \left(\begin{array}{@{}rr@{}} 0&\ \vec\sigma\\-\vec\sigma&0\end{array}\right)
\,,\end{eqnarray*}
it can be evaluated in closed form with \cite{BjD64}%
\begin{eqnarray} 
   u(k_1,s_1) = \sqrt{\frac{E_1+m_1}{2m_1}} \left( 
   \begin{array}{@{}r@{}r@{}}\chi_{s_1} \\ \displaystyle
   \phantom{-}\frac{\vec\sigma\cdot\vec k}{E_1(k)+m_1}\ \chi_{s_1} 
   \end{array} \right) 
\,,\nonumber\\
   u(k_2,s_2) = \sqrt{\frac{E_2+m_2}{2m_2}} \left( 
   \begin{array}{@{}r@{}r@{}}\chi_{s_2} \\ \displaystyle
   -\frac{\vec\sigma\cdot\vec k}{E_2(k)+m_2}\ \chi_{s_2} 
   \end{array} \right) 
\,.\label{eq:22}\end{eqnarray} 
The Pauli spinors are
$\chi_{\uparrow}  =\left(\begin{array}{@{}r@{}r@{}}1\\0\end{array}\right)$ and
$\chi_{\downarrow}=\left(\begin{array}{@{}r@{}r@{}}0\\1\end{array}\right)$.

\noindent
The components of the four current $j^\mu=(\rho,\vec j)$ are then 
\begin{eqnarray*}
\lefteqn{\hspace{-2em}
   \rho_1=\left[\overline u(k_1,s_1)\gamma^0 u(k'_1,s'_1)\right] =  \textstyle
   \sqrt{\frac{E_1(k)+m_1}{2m_1}\frac{E_1(k')+m_1}{2m_1}} 
}\nonumber\\ &\times&  \textstyle
   \Big\langle s_1\Big\vert 1 + 
   \frac{(\vec\sigma_1\cdot\vec k )}{(E_1(k )+m_1)}
   \frac{(\vec\sigma_1\cdot\vec k')}{(E_1(k')+m_1)} 
   \Big\vert s_1'\Big\rangle 
\,,\\ 
\lefteqn{\hspace{-2em}
   \vec j_1=\left[\overline u(k_1,s_1)\vec\gamma \phantom{^0}
   u(k'_1,s'_1)\right] =  \textstyle
   \sqrt{\frac{E_1(k)+m_1}{2m_1}\frac{E_1(k')+m_1}{2m_1}} 
}\nonumber\\ &\times&  \textstyle
   \Big\langle s_1\Big\vert 
   \frac{(\vec\sigma_1\cdot\vec k )\vec\sigma_1}{(E_1(k )+m_1)} +
   \frac{\vec\sigma_1(\vec\sigma_1\cdot\vec k')}{(E_1(k')+m_1)} 
   \Big\vert s_1'\Big\rangle 
\,.\end{eqnarray*} 
The spinor factor becomes then most straightforwardly:
\begin{eqnarray} 
\lefteqn{\hspace{-2ex}
   \langle s_1,s_2|S|s'_1,s'_2\rangle = \textstyle 
   \sqrt{\frac{E_1(k )+m_1}{2m_1}\frac{E_2(k )+m_2}{2m_2}} 
}\label{eq:28}\\ && \hspace{11ex}\times\textstyle 
   \sqrt{\frac{E_1(k')+m_1}{2m_1}\frac{E_2(k')+m_2}{2m_2}} 
   \hspace{1ex}\Big\langle s_1,s_2\Big|
\nonumber\\ &\times& \textstyle 
   \Big[\hspace{2ex}\left(1 + 
   \frac{(\vec\sigma_1\cdot\vec k)(\vec\sigma_1\cdot\vec k\,')}
   {(E_1(k)+m_1)(E_1(k')+m_1)}\right)
\nonumber\\ & & \textstyle \hspace{1ex} \times
   \left(1 + 
   \frac{(\vec\sigma_2\cdot\vec k) (\vec\sigma_2\cdot\vec k\,')}
   {(E_2(k)+m_2)(E_2(k')+m_2)}\right)
\nonumber\\ & & \textstyle \hspace{0ex} + \hspace{2ex}\Big(
   \frac{(\vec\sigma_1\cdot\vec k )\vec\sigma_1}{(E_1(k )+m_1)} +
   \frac{\vec\sigma_1(\vec\sigma_1\cdot\vec k')}{(E_1(k')+m_1)} \Big)
\nonumber\\ & & \textstyle \hspace{3ex} \cdot\Big(
   \frac{(\vec\sigma_2\cdot\vec k )\vec\sigma_2}{(E_2(k )+m_2)} +
   \frac{\vec\sigma_2(\vec\sigma_2\cdot\vec k')}{(E_2(k')+m_2)} \Big)
   \Big]\hspace{4ex}\Big|s'_1,s'_2\Big\rangle 
\,.\nonumber\end{eqnarray}
The expression is very much simpler than 
the long tables for the Lepage--Brodsky spinors \cite{Pau00c}, indeed.
The first two lines in the square bracket correspond 
to the product of the charges, $\rho_1\rho_2$, and the next two lines 
to the scalar product of the currents, $\vec j_1\vec j_2$.
Note that the plus sign of $\vec j_1\vec j_2$, as opposed
to the minus sign in
$\gamma^\mu\gamma_\mu= \gamma^0\gamma^0-\vec\gamma\vec\gamma$.
This is due to $\vec k_1=+\vec k$ and $\vec k_2=-\vec k$, 
see also Eq.(\ref{eq:22}).

\section{Conversion to a conventional Hamiltonian}
Thus far, the $4\times4$ matrix of the effective light-cone Hamiltonian 
in light-cone momentum representation, 
Eq.(\ref{eq:6}),  
\begin{eqnarray*} 
\lefteqn{\hspace{-4em}
   \int\!\!dk'_z d^2 \vec k'_{\!\perp}
   \langle h_1,h_2|H_\mathrm{eLC}
   (x,\vec k_{\!\perp};x',\vec k'_{\!\perp})|h'_1,h'_2\rangle 
}\nonumber\\ &\times&
   \psi_{h_1'h_2'}(k_z',\vec k_{\!\perp}') =
   M^2\psi_{h_1h_2}(x,\vec k_{\!\perp}) 
\,,\end{eqnarray*}
has been transformed to the $4\times4$ matrix 
of the effective light-cone Hamiltonian in usual momentum
representation, Eq.(\ref{eq:19}),  
\begin{eqnarray*} 
\lefteqn{\hspace{-28ex}
   \int\!\!d^3 \vec k'
   \langle s_1,s_2|H_\mathrm{eLC}
   (\vec k;\vec k')|s'_1,s'_2\rangle 
   \varphi_{s_1's_2'}(\vec k') =
   M^2\varphi_{s_1s_2}(\vec k) 
\,.}\end{eqnarray*}
Here and below the explicit summations over the
helicities are replaced by the Einstein convention. 
The spectrum of invariant mass-squared eigenvalues $M^2$
is unchanged by the transformation. 
One can subtract a c-number from an operator, 
and divide by a scale, and thus define
a new Hamiltonian $H$ with new eigenvalues $E$ 
but the same eigenfunctions $\varphi_{s_1s_2}$: 
\begin{eqnarray} 
\lefteqn{\hspace{-6em}
   H_\mathrm{eLC}(\vec k;\vec k')= \left(m_1+m_2\right)^2 + 
   2\left(m_1+m_2\right)H (\vec k;\vec k')
\,,}\nonumber\\  
   M^2 &=& \left(m_1+m_2\right)^2 + 
   2\left(m_1+m_2\right)E
\,.\end{eqnarray}
The eigenvalue $E$ has the dimension of an energy and is 
not to be confused with the single particle energy $E(k)$:
\begin{eqnarray} 
\lefteqn{\hspace{-9ex}
   \int\!\!d^3 \vec k'
   \langle s_1,s_2|H (\vec k;\vec k')|s'_1,s'_2\rangle 
   \varphi_{s_1's_2'}(\vec k') =
   E\varphi_{s_1s_2}(\vec k) 
\,,}\nonumber\\ \mbox{where} && \quad 
   H(\vec k;\vec k')= T(\vec k;\vec k')+U(\vec k;\vec k')
\,.\label{ieq:25}\end{eqnarray}
The kernels for kinetic and potential energy are given by
\begin{eqnarray} 
   T(\vec k;\vec k') &=& \phantom{-}
   \frac{k ^2}{2m_r}\delta^{(3)}(\vec k-\vec k') 
   \ \delta_{s_1s'_1}\delta_{s_2s'_2}\ B(k) 
\,,\label{ieq:23}\\ 
    U(\vec k;\vec k')&=& -\frac{\alpha_c(Q)}{2\pi^2 }
    \frac{S(\vec k;\vec k')}{Q^2}\ R(Q)\ \frac1{\sqrt{A(k)A(k')}}
\,,\label{eq:23}\end{eqnarray}
respectively.
Both $A(k)$ and $B(k)$ were defined in Eq.(\ref{neq:5}).

\section{Explicit calculation of the spinor factor}
This section is devoted to carry out explicitly the multiplications
in Eq.(\ref{eq:28}). 
The bilinear expressions with the same $\vec\sigma$'s 
can be simplified by means of the identities
\begin{eqnarray} 
   \begin{array} {rcrcl}
      (\vec k \cdot\vec\sigma) \vec\sigma &=& 
      \vec k  &+& i(\vec\sigma\wedge\vec k )
   \,,\\ 
      \vec\sigma (\vec k'\cdot\vec\sigma) &=& 
      \vec k\,' &-& i(\vec\sigma\wedge\vec k\,')
   \,,\\
      (\vec\sigma\cdot\vec k) (\vec\sigma\cdot\vec k\,') &=&
      \vec k\cdot\vec k\,' &+& i\vec\sigma\cdot(\vec k\wedge\vec k\,')
   \,.\end{array} 
\end{eqnarray} 
One gets thus identically in a first step:
\begin{eqnarray} 
\lefteqn{\hspace{-2ex}
   \langle s_1,s_2|S|s'_1,s'_2\rangle = \textstyle 
   \sqrt{\frac{E_1(k )+m_1}{2m_1}\frac{E_2(k )+m_2}{2m_2}} 
}\label{eq:28a}\\ && \hspace{11ex}\times\textstyle 
   \sqrt{\frac{E_1(k')+m_1}{2m_1}\frac{E_2(k')+m_2}{2m_2}} 
   \hspace{1ex}\Big\langle s_1,s_2\Big|
\nonumber\\ &\times& \textstyle 
   \Big[\hspace{2ex}\left(1 + 
   \frac{\vec k\cdot\vec k\,' + i\vec\sigma_1\cdot\vec k\wedge\vec k\,'}
   {(E_1(k)+m_1)(E_1(k')+m_1)}\right)
\nonumber\\ & & \textstyle \hspace{1ex} \times
   \left(1 + 
   \frac{\vec k\cdot\vec k\,' + i\vec\sigma_2\cdot\vec k\wedge\vec k\,'}
   {(E_2(k)+m_2)(E_2(k')+m_2)}\right)
\nonumber\\ & & \textstyle \hspace{0ex} + \hspace{2ex}\Big(
   \frac{\vec k    + i\vec\sigma_1\wedge\vec k   }{(E_1(k )+m_1)} +
   \frac{\vec k\,' - i\vec\sigma_1\wedge\vec k\,'}{(E_1(k')+m_1)} \Big)
\nonumber\\ & & \textstyle \hspace{3ex} \cdot\Big(
   \frac{\vec k    + i\vec\sigma_2\wedge\vec k   }{(E_2(k )+m_2)} +
   \frac{\vec k\,' - i\vec\sigma_2\wedge\vec k\,'}{(E_2(k')+m_2)} \Big)
   \Big]\hspace{4ex}\Big|s'_1,s'_2\Big\rangle 
\,.\nonumber\end{eqnarray}
Carrying out the multiplications one gets in a second step:
\begin{eqnarray} 
\lefteqn{\hspace{-2ex}
   \langle s_1,s_2|S|s'_1,s'_2\rangle = \textstyle 
   \sqrt{\frac{E_1(k )+m_1}{2m_1}\frac{E_2(k )+m_2}{2m_2}} 
}\label{eq:28c}\\ && \hspace{11ex}\times\textstyle 
   \sqrt{\frac{E_1(k')+m_1}{2m_1}\frac{E_2(k')+m_2}{2m_2}} 
   \hspace{1ex}\Big\langle s_1,s_2\Big|\Big[1+
\nonumber\\ &+&  \textstyle 
   \frac{\vec k\cdot\vec k\,' + i\vec\sigma_1\cdot\vec k\wedge\vec k\,'}
   {(E_1(k)+m_1)(E_1(k')+m_1)} +
   \frac{\vec k\cdot\vec k\,' + i\vec\sigma_2\cdot\vec k\wedge\vec k\,'}
   {(E_2(k)+m_2)(E_2(k')+m_2)} 
\nonumber\\ \textstyle &+&  \textstyle 
   \frac{\vec k\cdot\vec k\,' + i\vec\sigma_1\cdot\vec k\wedge\vec k\,'}
   {(E_1(k)+m_1)(E_1(k')+m_1)}  
   \frac{\vec k\cdot\vec k\,' + i\vec\sigma_2\cdot\vec k\wedge\vec k\,'}
   {(E_2(k)+m_2)(E_2(k')+m_2)} 
\nonumber\\ \textstyle &+&  \textstyle 
   \frac{\vec k    + i\vec\sigma_1\wedge\vec k   }{(E_1(k )+m_1)} \cdot   
   \frac{\vec k    + i\vec\sigma_2\wedge\vec k   }{(E_2(k )+m_2)} +
   \frac{\vec k\,' - i\vec\sigma_1\wedge\vec k\,'}{(E_1(k')+m_1)} \cdot 
   \frac{\vec k\,' - i\vec\sigma_2\wedge\vec k\,'}{(E_2(k')+m_2)} 
\nonumber\\ \textstyle &+&  \textstyle 
   \frac{\vec k    + i\vec\sigma_1\wedge\vec k   }{(E_1(k )+m_1)} \cdot 
   \frac{\vec k\,' - i\vec\sigma_2\wedge\vec k\,'}{(E_2(k')+m_2)}  
\nonumber\\ &+& \textstyle 
   \frac{\vec k\,' - i\vec\sigma_1\wedge\vec k\,'}{(E_1(k')+m_1)} \cdot 
   \frac{\vec k    + i\vec\sigma_2\wedge\vec k   }{(E_2(k )+m_2)}    
   \hspace{7em}\Big]\Big|s'_1,s'_2\Big\rangle 
\,.\nonumber\end{eqnarray}
With Eqs.(\ref{eq:29a},\ref{eq:29b},\ref{eq:29c}) below, 
one gets in a third step:
\begin{eqnarray} 
\lefteqn{\hspace{-0ex}
   \langle s_1,s_2|S|s'_1,s'_2\rangle = \textstyle 
   \sqrt{\frac{E_1(k )+m_1}{2m_1}\frac{E_2(k )+m_2}{2m_2}} 
}\label{eq:28d}\\ && \hspace{11ex}\times\textstyle 
   \sqrt{\frac{E_1(k')+m_1}{2m_1}\frac{E_2(k')+m_2}{2m_2}} 
   \hspace{1ex}\Big\langle s_1,s_2\Big|\Big[1+
\nonumber\\ &+&  \textstyle 
   \frac{\vec k\cdot\vec k\,' + i\vec\sigma_1\cdot\vec k\wedge\vec k\,'}
   {(E_1(k)+m_1)(E_1(k')+m_1)} +
   \frac{\vec k\cdot\vec k\,' + i\vec\sigma_2\cdot\vec k\wedge\vec k\,'}
   {(E_2(k)+m_2)(E_2(k')+m_2)} 
\nonumber\\ \textstyle &+&  \textstyle 
   \frac{\vec k^2 \vec k'\,^2 - \left(\vec k\wedge\vec k\,'\right)^2 -  
   \left(\vec\sigma_1\cdot\vec k\wedge\vec k\,'\right)
   \left(\vec\sigma_2\cdot\vec k\wedge\vec k\,'\right) +i
   \left(\vec\sigma_1+\vec\sigma_2\right)\cdot\vec k\wedge\vec k\,'
   \left(\vec k\cdot\vec k\,'\right)}
   {(E_1(k)+m_1)(E_1(k')+m_1)(E_2(k)+m_2)(E_2(k')+m_2)}  
\nonumber\\ \textstyle &+&  \textstyle 
   \frac{\vec k ^2 - 
   \left(\vec\sigma_1\wedge\vec k   \right) \cdot 
   \left(\vec\sigma_2\wedge\vec k   \right)}
   {(E_1(k )+m_1)(E_2(k )+m_2)} +
   \frac{\vec k '\,^2 - 
   \left(\vec\sigma_1\wedge\vec k\,'\right) \cdot 
   \left(\vec\sigma_2\wedge\vec k\,'\right)}
   {(E_1(k')+m_1)(E_2(k')+m_2)}  
\nonumber\\ \textstyle &+&  \textstyle 
   \frac{\vec k\vec k\,' + 
   \left(\vec\sigma_1\wedge\vec k   \right) \cdot 
   \left(\vec\sigma_2\wedge\vec k\,'\right) + i
   \left(\vec\sigma_1+\vec\sigma_2\right)\cdot
   \left(\vec k \wedge\vec k '\right)}
   {(E_1(k )+m_1)(E_2(k')+m_2)}  
\nonumber\\ &+& \textstyle 
   \frac{\vec k\vec k\,' + 
   \left(\vec\sigma_1\wedge\vec k\,'\right) \cdot 
   \left(\vec\sigma_2\wedge\vec k   \right) + i
   \left(\vec\sigma_1+\vec\sigma_2\right)\cdot
   \left(\vec k '\wedge\vec k \right)}
   {(E_1(k')+m_1)(E_2(k )+m_2)}
   \hspace{2em} \Big]\Big|s'_1,s'_2\Big\rangle 
\,.\nonumber\end{eqnarray}
Here, the familiar vector identities like
\begin{eqnarray} 
   \begin{array} {@{}r@{}c @{}c@{}c@{}c @{}lclcl}
       &\vec a\wedge\vec b& &\cdot & &\vec c &=& \vec a\cdot\vec b\wedge\vec c
   \,,\\
      (&\vec a\wedge\vec b&)&\wedge& &\vec c &=& 
      (\vec a\cdot\vec c)\vec b &-& 
      (\vec b\cdot\vec c)\vec a 
   \,,\\
      (&\vec a\wedge\vec b&)&\cdot &(&\vec c \wedge\vec d) &=&
      (\vec a\cdot\vec c) \vec b\cdot\vec d &-& 
      (\vec b\cdot\vec c) \vec a\cdot\vec d
   \,,\end{array} 
\label{eq:29a}\end{eqnarray} 
were used to derive particularly 
\begin{eqnarray} && 
   \left(\vec k\cdot \vec k\,'\right)^2 = \vec k\,^2 \vec k\,'\,^2 -
   \left(\vec k\wedge\vec k\,'\right)^2
\,,\label{eq:29b}\\  && 
   \left(\vec k    + i\vec\sigma_1\wedge\vec k   \right) \cdot 
   \left(\vec k\,' - i\vec\sigma_2\wedge\vec k\,'\right) =
\label{eq:29c}\\  && 
   \vec k  \vec k\,' +
   (\vec\sigma_1\wedge\vec k )\cdot
   (\vec\sigma_2\wedge\vec k\,') +
   i\left(\vec\sigma_1+\vec\sigma_2\right)\cdot
   \left(\vec k \wedge\vec k\,'\right)
\,.\nonumber\end{eqnarray} 
In deriving Eq.(\ref{eq:28d}), Eq.(\ref{eq:29b}) was used 
in line 3 of the square bracket of the equation.
In lines 5 and 6, Eq.(\ref{eq:29c}) was applied.
In line 4, Eq.(\ref{eq:29c}) was applied as well, but considering
the fact that the wedge product disappears for $\vec k =\vec k '$.

Eq.(\ref{eq:28d}) emphasizes the wedge product $\vec k\wedge\vec k\,'$.
As to be seen below, it is closely related to the 
orbital angular momentum $\vec L$, which likes for example
to be dotted into the spin to produce  
spin orbit coupling $\vec L\cdot\vec\sigma$. 
It is therefore reasonable to divide the spinor factor into two parts, 
\begin{eqnarray} \textstyle 
   S=S_0+S_1 
\,,\end{eqnarray} 
with $S_0$ being independent of the wedge product, 
and with $S_1$ depending \emph{only} on the wedge product. Thus:
\begin{eqnarray} 
\lefteqn{\hspace{-0ex}
   \langle s_1,s_2|S_0|s'_1,s'_2\rangle = \textstyle 
   \sqrt{\frac{E_1(k )+m_1}{2m_1}\frac{E_2(k )+m_2}{2m_2}} 
}\label{eq:28e}\\ && \hspace{14ex}\times\textstyle 
   \sqrt{\frac{E_1(k')+m_1}{2m_1}\frac{E_2(k')+m_2}{2m_2}} 
   \hspace{1ex}\big\langle s_1,s_2\big|\big[1+
\nonumber\\ &+&  \textstyle 
   \frac{\vec k\cdot\vec k\,'}
   {(E_1(k)+m_1)(E_1(k')+m_1)} +
   \frac{\vec k\cdot\vec k\,'}
   {(E_2(k)+m_2)(E_2(k')+m_2)} 
\nonumber\\ \textstyle &+&  \textstyle 
   \frac{\vec k^2 \vec k'\,^2}
   {(E_1(k)+m_1)(E_1(k')+m_1)(E_2(k)+m_2)(E_2(k')+m_2)}  
\nonumber\\ \textstyle &+&  \textstyle 
   \frac{\vec k ^2 - 
   \left(\vec\sigma_1\wedge\vec k   \right) \cdot 
   \left(\vec\sigma_2\wedge\vec k   \right)}
   {(E_1(k )+m_1)(E_2(k )+m_2)} +
   \frac{\vec k '\,^2 - 
   \left(\vec\sigma_1\wedge\vec k\,'\right) \cdot 
   \left(\vec\sigma_2\wedge\vec k\,'\right)}
   {(E_1(k')+m_1)(E_2(k')+m_2)}  
\nonumber\\ \textstyle &+&  \textstyle 
   \frac{\vec k\vec k\,' + 
   \left(\vec\sigma_1\wedge\vec k   \right) \cdot 
   \left(\vec\sigma_2\wedge\vec k\,'\right)}
   {(E_1(k )+m_1)(E_2(k')+m_2)} + 
   \frac{\vec k\vec k\,' + 
   \left(\vec\sigma_1\wedge\vec k   \right) \cdot 
   \left(\vec\sigma_2\wedge\vec k\,'\right)}
   {(E_1(k')+m_1)(E_2(k )+m_2)}
   \big]\big|s'_1,s'_2\big\rangle 
,\nonumber\end{eqnarray}
and 
\begin{eqnarray} 
\lefteqn{\hspace{-0ex}
   \langle s_1,s_2|S_1|s'_1,s'_2\rangle = \textstyle 
   \sqrt{\frac{E_1(k )+m_1}{2m_1}\frac{E_2(k )+m_2}{2m_2}} 
}\label{eq:28f}\\ && \hspace{14ex}\times\textstyle 
   \sqrt{\frac{E_1(k')+m_1}{2m_1}\frac{E_2(k')+m_2}{2m_2}} 
   \hspace{5ex}\big\langle s_1,s_2\big|\big[ 
\nonumber\\ & &  \textstyle 
   \frac{i\vec\sigma_1\cdot\vec k\wedge\vec k\,'}
   {(E_1(k)+m_1)(E_1(k')+m_1)} +
   \frac{i\vec\sigma_2\cdot\vec k\wedge\vec k\,'}
   {(E_2(k)+m_2)(E_2(k')+m_2)} 
\nonumber\\ \textstyle &+&  \textstyle 
   \frac{- \left(\vec k\wedge\vec k\,'\right)^2 -  
   \left(\vec\sigma_1\cdot\vec k\wedge\vec k\,'\right)
   \left(\vec\sigma_2\cdot\vec k\wedge\vec k\,'\right) +i
   \left(\vec\sigma_1+\vec\sigma_2\right)\cdot\vec k\wedge\vec k\,'
   \left(\vec k\cdot\vec k\,'\right)}
   {(E_1(k)+m_1)(E_1(k')+m_1)(E_2(k)+m_2)(E_2(k')+m_2)}  
\nonumber\\ \textstyle &+&  \textstyle 
   \frac{i
   \left(\vec\sigma_1+\vec\sigma_2\right)\cdot
   \left(\vec k \wedge\vec k\,'\right)}
   {(E_1(k )+m_1)(E_2(k')+m_2)} + 
   \frac{i
   \left(\vec\sigma_1+\vec\sigma_2\right)\cdot
   \left(\vec k \wedge\vec k\,'\right)}
   {(E_1(k')+m_1)(E_2(k )+m_2)}
   \big]\big|s'_1,s'_2\big\rangle 
.\nonumber\end{eqnarray}
One should note that $S_1$ can be omitted if one is interested
only in a spherically symmetric wave function.

\section{The Fourier approximation}
Of course, one can diagonalize the Hamiltonian 
directly in momentum space, see Eq.(\ref{eq:23}). 
But momentum space does not appeal to
intuition, at least not to my intuition.
One would like to Fourier transform to configuration space,
in order to interpret and to understand what one is doing.
This however turns out to be impossible for the general case,
for two reasons: (1) The Fourier transform can not be taken
analytically, and (2) the square root in $E(k)=\sqrt{m^2+k^2}$ 
induces strong non-localities in the potential energy part 
of the Hamiltonian, see for example App.~\ref{asec:6}. 

In the further analysis, 
I will apply therefore what I call the \emph{Fourier approximation} 
\begin{eqnarray} \textstyle
   E(k) \Longrightarrow m
\,.\label{eq:26}\end{eqnarray}
But one must be careful. 
The un-considered application of Eq.(\ref{eq:26}) for example
to the l.h.s. of Eq.(\ref{eq:19}) gives plain non-sense.

The non-sense is avoided by substituting in Eq.(\ref{eq:23})
\begin{eqnarray}
\begin{array}{rcl}
   A(k)&\Longrightarrow& 1 
\,,\\
   B(k)&\Longrightarrow& 1
\,.\end{array}
\label{eq:a26}\end{eqnarray}
Similarly, one substitutes
the four-momentum transfers in Eq.(\ref{neq:2}) according to
\begin{eqnarray}
\begin{array}{rcl}
   Q_{q}^2      &=& (\vec k-\vec k')^2 -(E_1(k)-E_1(k'))^2 
   \Longrightarrow  (\vec k-\vec k')^2
\,,\\
   Q_{\bar q}^2 &=& (\vec k-\vec k')^2 -(E_2(k)-E_2(k'))^2 
   \Longrightarrow  (\vec k-\vec k')^2
\,.\end{array}
\end{eqnarray}
I try to avoid the expression ``non-relativistic approximation.''
Relativistic or non-relativistic motion is usually associated with 
the kinetic energy. But here the Fourier approximation takes
place essentially in the interaction, under the integral, 
and is applied, as mentioned, 
for the sole purpose of making a Fourier transform possible.
Whether this is a good approximation or not, 
whether Eq.(\ref{eq:a26}) is applicable or not,  
can be answered only in the solution, \emph{a posteriori}.
In any case, this question must remain on the agenda.

In the Fourier approximation, 
Eqs.(\ref{eq:28e},\ref{eq:28f}) become very much more simplified.
Suppressing explicit reference to the bras and kets $|s_1,s_2\rangle $
one gets simply
\begin{eqnarray} 
\lefteqn{\hspace{-0ex}
   S_0 = \textstyle
   1+ 
   \frac{\vec k ^2 - 
   \left(\vec\sigma_1\wedge\vec k   \right) \cdot 
   \left(\vec\sigma_2\wedge\vec k   \right)}{4m_1 m_2} +
   \frac{\vec k '\,^2 - 
   \left(\vec\sigma_1\wedge\vec k\,'\right) \cdot 
   \left(\vec\sigma_2\wedge\vec k\,'\right)}{4m_1 m_2}  
}\label{eq:28g}\\ &+&  \textstyle 
   \frac{\vec k\vec k\,' + 
   \left(\vec\sigma_1\wedge\vec k   \right) \cdot 
   \left(\vec\sigma_2\wedge\vec k\,'\right)}
   {2 m_1 m_2} +
   \frac{\vec k\cdot\vec k\,'}{4m_1^2} +
   \frac{\vec k\cdot\vec k\,'}{4m_2^2} + 
   \frac{1}{16}\frac{\vec k^2}{m_1^2}\frac{\vec k'\,^2}{m_2^2}  
,\nonumber\\
\lefteqn{\hspace{-0ex}
   S_1 = \textstyle 
   \frac{i\vec\sigma_1\cdot\vec k\wedge\vec k\,'} {4 m_1^2} +
   \frac{i\vec\sigma_2\cdot\vec k\wedge\vec k\,'} {4 m_2^2} +
   \frac{i(\vec\sigma_1+\vec\sigma_2)\cdot\vec k\wedge\vec k\,'}{2 m_1 m_2 }
}\label{eq:28h}\\ &+&  \textstyle
   \frac{\vec k\cdot\vec k '}{4 m_1 m_2}\frac{i\left(
   \vec\sigma_1+\vec\sigma_2\right)\cdot\vec k\wedge\vec k '}{4 m_1 m_2} -
   \frac{\left(\vec k\wedge\vec k\,'\right)^2  +
   \left(\vec\sigma_1\cdot\vec k\wedge\vec k\,'\right)
   \left(\vec\sigma_2\cdot\vec k\wedge\vec k\,'\right)}{(4 m_1 m_2)^2} 
,\nonumber\end{eqnarray}
respectively. 

With the usual hyperfine approximation, see for example \cite{BjD64},
\begin{eqnarray} 
   (\vec\sigma_1\wedge\vec b)\cdot(\vec\sigma_2\wedge\vec c)
   &\Longrightarrow& \phantom{-} \textstyle
   \frac23(\vec\sigma_1\vec\sigma_2)(\vec b\vec c)
\,,\label{eq:45}\end{eqnarray} 
with $\vec b$ or $\vec c$ being any of the vectors 
$\vec k$, $\vec k'$, or $\vec k\wedge\vec k'$,
these equations can be simplified even more. One gets
\begin{eqnarray} \textstyle 
   S_0 &=& 1  +
   \underbrace{\textstyle
   (-\vec\sigma_1\vec\sigma_2 )\frac{(\vec k-\vec k ')^2}{6m_1 m_2}
   }_{\mathrm{hyperfine\ term}} +
   \underbrace{\textstyle
   \frac{(\vec k+\vec k ')^2}{4m_1 m_2}
   }_{\mathrm{kinetic}} +
\nonumber\\ && \phantom{1} + 
   \underbrace{\textstyle
   \frac{\vec k\vec k'}{4m_1m_2}\Big[\frac{m_1}{m_2} + \frac{m_2}{m_1}\Big]
   }_{\mathrm{Darwin}} + 
   \underbrace{\textstyle
   \frac{1}{16}\frac{\vec k^2}{m_1^2}\frac{\vec k'\,^2}{m_2^2}  
   }_{= 0} 
\,,\\ &=&
   1+S_\mathrm{hf}+S_\mathrm{K}+S_\mathrm{D}
\,.\label{ieq:43}\end{eqnarray} 
The last term must be suppressed for 
consistency with Eq.(\ref{eq:26}). 
Obviously, the identity
\begin{eqnarray*}  
   \frac{1}{m_1^2} + \frac{1}{m_2^2} =
   \frac{1}{m_1m_2}\Big[\frac{m_1}{m_2}+\frac{m_2}{m_1}\Big]
\end{eqnarray*}
has been applied here. 
Correspondingly one gets 
\begin{eqnarray} \textstyle 
   S_1 &=& \textstyle
   \underbrace{\textstyle
   \frac{i\vec\sigma_1\cdot\vec k\wedge\vec k '} {4m_1} 
   \Big[\frac{1}{m_r }+\frac{1}{m_1}\Big]
   }_{\mathrm{spin-orbit\ term}_1} +
   \underbrace{\textstyle
   \frac{i\vec\sigma_2\cdot\vec k\wedge\vec k '} {4m_2} 
   \Big[\frac{1}{m_r }+\frac{1}{m_2}\Big] 
   }_{\mathrm{spin-orbit\ term}_2} 
\\ &+& \textstyle
   \underbrace{\textstyle
   \frac{-\left(\vec k\wedge\vec k'\right)^2}{(4 m_1 m_2)^2}   
   }_{\mathrm{L}^2\mathrm{-term}} +
   \underbrace{\textstyle
   \frac{-\vec\sigma_1\vec\sigma_2}{6 m_1 m_2}
   \frac{\left(\vec k\wedge\vec k'\right)^2}{4 m_1 m_2}   
   }_{\mathrm{hyperfine-L}^2\mathrm{-term}} +\textstyle 
   \underbrace{\textstyle
   \frac{i\left(
   \vec\sigma_1+\vec\sigma_2\right)\cdot\vec k\wedge\vec k '}{4 m_1 m_2}
   \frac{\vec k\cdot\vec k '}{4 m_1 m_2} 
   }_{\mathrm{spin-orbit\ Darwin\ term}} 
\nonumber\\ &=& \textstyle
   S_\mathrm{so_1}+S_\mathrm{so_2}+S_\mathrm{L}+S_\mathrm{hfL}+S_\mathrm{soD} 
.\label{ieq:45}\end{eqnarray}
The nomenclature is more transparent in configuration space.
The kernel of the potential energy is then
\begin{eqnarray} 
    U(\vec k;\vec k')&=& -\frac{\alpha_c(q)}{2\pi^2 }
    \frac{R(q)}{q^2}\ S(\vec k;\vec k')
\,,\label{ieq:46}\end{eqnarray}
with $\vec q = \vec k-\vec k'$ and $S=S_0+S_1$.

\subsection{The hyperfine approximation}
Let me first investigate  
\begin{eqnarray} 
   (\vec \sigma\wedge\vec b)\wedge\vec c =
   (\vec \sigma\cdot\vec c)\vec b - 
   (\vec b\cdot\vec c)\vec \sigma 
\,.\label{eq:44}\end{eqnarray} 
The 3-component of Eq.(\ref{eq:44}) is then
\begin{eqnarray*} 
   \left[
   (\vec \sigma\wedge\vec b)\wedge\vec c\right]_3 &=& 
   (\sigma_1 c_1+\sigma_2 c_2+\sigma_3 c_3)b_3   
\\ &-&
   (b_1 c_1+b_2 c_2+b_3 c_3)\sigma_3 
\nonumber\\ &=&
   (\sigma_1 c_1+\sigma_2 c_2)b_3 - 
   (b_1 c_1+b_2 c_2)\sigma_3 
\,.\end{eqnarray*} 
If one has reasons to believe that the off-diagonal components
of $b_i c_j$ cancel by symmetry considerations, 
\textit{i.e.} 
$\langle b_i c_j\rangle \simeq \frac13\delta_{ij}\vec b\vec c$,
one gets 
$\langle (\vec \sigma\wedge\vec b)\wedge\vec c\rangle_3
 \simeq -\frac23\sigma_3 (\vec b\vec c) $.
The other components behave correspondingly. 

Since $(\vec\sigma_1\wedge\vec b)\cdot(\vec\sigma_2\wedge\vec c)=
-\left[(\vec\sigma_1\wedge\vec b)\wedge\vec c\right]\cdot\vec\sigma_2$,
Eq.(\ref{eq:45}) is valid approximately.   
One is accustomed to this substitution from the theory of
hyperfine interactions \cite{BjD64}, and I will refer to it as the 
\emph{the hyperfine approximation}. As shown in the appendices,
the hyperfine approximation is a rather weak assumption. 
Sometimes it is even exact.

\subsection{Deriving the Singlet-Triplet model}
The wedge product 
$(\vec\sigma_1\wedge\vec k)\cdot(\vec\sigma_2\wedge\vec k')$ 
is defined in Eq.(\ref{eq:29a}).
Replacing it by
\begin{eqnarray} 
   \left(\vec\sigma_1\wedge\vec k  \right) \cdot 
   \left(\vec\sigma_2\wedge\vec k '\right) 
   \Longrightarrow 
   \left(\vec\sigma_1\vec\sigma_2\right)  
   \left(\vec k\vec k '\right) 
\,,\label{eq:39}\end{eqnarray}
rather than by the hyperfine approximation in Eq.(\ref{eq:45}),  
one gets from Eq.(\ref{eq:28g}) in a first step: 
\begin{eqnarray} \textstyle 
   S_0 = \textstyle 
   1&+&\textstyle\frac{\vec k\vec k\,'}{4m_1^2} +
   \frac{\vec k\vec k\,'}{4m_2^2}   
\nonumber\\ &+&  \textstyle 
   \frac{\vec k   ^2(1-\vec\sigma_1\vec\sigma_2) +
         \vec k'\,^2(1-\vec\sigma_1\vec\sigma_2) + 2
         \vec k\vec k'(1+\vec\sigma_1\vec\sigma_2)}{4m_1 m_2} 
\,.\nonumber\end{eqnarray} 
Adding and subtracting $2\vec k\vec k\,'(1-\vec\sigma_1\vec\sigma_2)$ 
gives 
\begin{eqnarray} \textstyle 
   S_0 &=& \textstyle
   1+(1-\vec\sigma_1\vec\sigma_2)\frac{(\vec k-\vec k')^2}{4m_1 m_2} +
   \frac{\vec k\vec k'}{4}
   \Big[\frac{1}{m_1^2} +\frac{1}{m_2^2} + \frac{4}{m_1m_2} \Big] 
\nonumber\\ &=&  \textstyle 
   1+(1-\vec\sigma_1\vec\sigma_2)\frac{Q^2}{4m_1 m_2} +
   \frac{\vec k\vec k'}{4}
   \Big[\frac{1}{m_r^2}+\frac{2}{m_1m_2}\Big] 
\,.\label{eq:42}\end{eqnarray} 
In previous work \cite{Pau03a}, the \emph{matrix} $S$ was replaced by
its \emph{diagonal elements},
\begin{eqnarray*} 
    \langle s_1,s_2\vert S\vert s_1',s_2'\rangle 
    &\Longrightarrow&
    \delta_{s_1,s_1'}\;\delta_{s_2,s_2'}\;
    \langle s_1,s_2\vert S\vert s_1,s_2\rangle
\,.\end{eqnarray*} 
Eq.(\ref{eq:42}) gives then,
up to terms proportional to $\vec k\vec k'$,
\begin{eqnarray*} 
    \frac{\langle s_1,s_2\vert S\vert s_1,s_2\rangle}{Q^2}
    &=&
    \left\{
    \begin{array}{ll}
    \frac1{Q^2}+\frac2{4 m_1 m_2}, 
      &\mbox{ for $s_1 = - s_2 $,}\\
    \frac1{Q^2},\phantom{+\frac2{4 m_1 m_2}} 
      &\mbox{ for $s_1 = \phantom{-} s_2 $.}\\
    \end{array}\right.
\end{eqnarray*} 
One has thus \emph{derived the Singlet-Triplet model} \cite{Pau03a} 
without \textit{ad hoc} procedures.
The model has played an important role in the development of the theory,
particularly its renormalization.

\section{The Hamiltonian in configuration space}
The eigenvalue equation (\ref{ieq:25}) in momentum space, 
\begin{eqnarray*} 
\lefteqn{\hspace{-12em}
   \int\!\!d^3 \vec k'
   \langle s_1,s_2|H (\vec k;\vec k')|s'_1,s'_2\rangle 
   \varphi_{s_1's_2'}(\vec k') =
   E\varphi_{s_1s_2}(\vec k) 
\,,}\end{eqnarray*}
can be Fourier transformed to configuration space, 
\begin{eqnarray*} 
\lefteqn{\hspace{-12em}
   \int\!\!d^3 \vec r'
   \langle s_1,s_2|H (\vec r;\vec r')|s'_1,s'_2\rangle 
   \Psi_{s_1's_2'}(\vec r') =
   E\Psi_{s_1s_2}(\vec r) 
\,.}\end{eqnarray*}
The eigenvalue equation
\begin{eqnarray}
   H\Psi  = E\Psi
\,,\label{eq:50}\end{eqnarray}
has then the structure of the familiar Schr\"odinger equation.
Quite in general, working in configuration space has the advantage
of being closer to conventional quantum mechanics and to
phenomenological models where our intuition comes from. 

Fourier transformations need a sign convention, 
\begin{eqnarray} 
   H (\vec r;\vec r') &=&
   \int\! d^3\!\vec k\ \mathrm{e}^{i\vec k\vec r}
   \int\! \frac{d^3\!\vec k'}{(2\pi)^3} 
   \mathrm{e}^{-i\vec k'\vec r'}
   H (\vec k;\vec k')
\,,\label{eq:47}\\  
   \Psi_{s_1s_2}(\vec r) &=&
   \int\! d^3\!\vec k\ \mathrm{e}^{i\vec k\vec r}
   \varphi_{s_1s_2}(\vec k) 
\,.\end{eqnarray}
It could be a source of endless confusion that
$H (\vec k;\vec k')$ and $H (\vec r;\vec r')$ are denoted by
the same symbol $H$. But this should not be a problem 
since the meaning is usually clear from the context.

Following the nomenclature of Eqs.(\ref{ieq:43},\ref{ieq:45})
the Hamiltonian is decomposed into
\begin{eqnarray}
   H   &=& T + V_0 + V_1 
\,,\qquad\qquad\mbox{ with }\\ 
   V_0 &=& V\phantom{_\mathrm{so}} + V_\mathrm{hf} + 
           V_\mathrm{K} + V_\mathrm{D}
\,,\\ 
   V_1 &=& V_\mathrm{so_1} + V_\mathrm{so_2} + V_\mathrm{L} + 
           V_\mathrm{hfL}  + V_\mathrm{soD} 
\,.\end{eqnarray}
The kinetic and the potential energy are respectively
\begin{eqnarray}
   T &=& \frac{\vec p\,^2}{2m_r}  
\,,\\
   V &=& 
   -\int\!\!d^3\!\vec q\ \mathrm{e}^{i\vec q\vec r}
   \frac{\alpha_c(q)}{2\pi^2}\frac{R(q)}{q^2} 
\,.\end{eqnarray}
The hyperfine, kinetic, and Darwin potentials are
\begin{eqnarray}   
   V_\mathrm{hf} &=& \textstyle \phantom{-}
   \frac{\vec\sigma_1\vec\sigma_2}{6m_1 m_2} \nabla^2 V   
,\label{eq:57}\\ \textstyle
   V_\mathrm{K} &=& \textstyle \phantom{-} 
   \frac{V}{m_1 m_2} \vec p\,^2   
,\\ \textstyle
   V_\mathrm{D} &=& \textstyle -
   \left(\frac{\nabla^2V}{4m_1m_2}+\frac{V\vec p\,^2}{16m_1m_2}\right)
   \left[\frac{m_1}{m_2} + \frac{m_2}{m_1}\right] 
,\end{eqnarray} 
respectively. The spin-orbit, L$^2$, hyperfine-L$^2$, and
spin-orbit-Darwin potentials are
\begin{eqnarray}   
   V_\mathrm{so_1} &=& \textstyle  
   \frac{\vec\sigma_1\cdot\vec L} {4m_1} 
   \Big[\frac{1}{m_r }+\frac{1}{m_1}\Big]
   \Big[\frac{1}{r}\frac{dV}{dr}\Big] 
,\\ \textstyle
   V_\mathrm{L} &=& \textstyle 
   \frac{\vec L^2}{(4 m_1 m_2)^2} 
   \Big[\frac{1}{r^2}\frac{d^2V}{dr^2}\Big] 
,\\ \textstyle
   V_\mathrm{hfL} &=& \textstyle 
   \frac{\vec\sigma_1\vec\sigma_2}{6 m_1 m_2}
   \frac{\vec L^2}{4 m_1 m_2}\ %
   \Big[\frac{1}{r^2}\frac{d^2V}{dr^2}\Big] 
,\\ \textstyle
   V_\mathrm{soD} &=& \textstyle  
   \frac{\left(
   \vec\sigma_1+\vec\sigma_2\right)\cdot\vec L}
   {\left(4 m_1 m_2\right)^2}\ \Big(
   \Big[\frac{1}{4r}\frac{dV}{dr}\Big]\nabla^2 -
   \Big[\frac{1}{r}\frac{d(\nabla^2V)}{dr}\Big] \Big)
,\end{eqnarray}
respectively. They depend on the angular momentum operator
$\vec L= \vec r\wedge\vec p$,  
see App.~\ref{asec:DiffTe},
and vanish for spherically symmetric wave functions.
The linear momentum operator is denoted by $\vec p \equiv -i\vec\nabla$.

Some of the above terms appear also in the analysis of the 
Dirac equation with an external vector potential, 
see \textit{for example} \cite{BjD64}.
One should emphasize that the present analysis is 
more complete since retardation effects are fully included.
They generate the remainder of the terms in the above.

\section{Spherically symmetric wave functions}
It is our freedom to solve Eq.(\ref{eq:50}) only for wave functions 
with spherical symmetry, $\Psi_{s_1s_2}(\vec r)=\Psi_{s_1s_2}(r)$. 
The s-states, in a way, are also the most interesting, 
since practically all hadrons suitable for beams or targets 
have this symmetry. And for them one can solve the problem rigorously.

Since there is no angular momentum, one can couple 
the quark Bj\o rken-Drell spins into total spin and thus 
total angular momentum, by means of Wigner-Eckart \cite{Edm64},
\begin{eqnarray} \textstyle
   \vert S,S_z\rangle &=& \sum_{s_1,s_2} \textstyle
   \langle \frac12 s_1 \frac12 s_2\vert S,S_z\rangle
   \vert s_1\rangle \vert s_2\rangle 
\,,\end{eqnarray} 
with the total spin (or total angular momentum $J$) is
either $S=0$ for singlets or $S=1$ for triplets.
Using the completeness relations of the Clebsch-Gordan coefficients
\begin{eqnarray} \textstyle
   \sum_{S,S_z} \textstyle
   \langle \frac12 s_1 \frac12\ s_2\vert S,S_z\rangle
   \langle S,S_z\vert \frac12 s_1' \frac12\ s_2'\rangle &=& 
   \delta_{s_1,s_1'} \delta_{s_2,s_2'} 
\,,\end{eqnarray}
I can perform (for the last time) a unitary transformation 
\begin{eqnarray} \textstyle
   \widetilde \varphi_{S,S_z}(\vec k) &=&
   \sum_{s_1,s_2} \textstyle
   \varphi_{s_1s_2}(\vec k)  
   \langle \frac12 s_1 \frac12 s_2\vert S,S_z\rangle
\,,\\
   \widetilde \Psi_{S,S_z}(r) &=& 
   \sum_{s_1,s_2} \textstyle
   \Psi_{s_1s_2}(r) 
   \langle \frac12 s_1 \frac12 s_2\vert S,S_z\rangle
\,,\end{eqnarray} 
and solve the Hamiltonian equations
\begin{eqnarray}
   H\widetilde \Psi_{S,S_z}(r) &=& E\widetilde \Psi_{S,S_z}(r)
\,,\\
   \mbox{or}\qquad  
   H\otimes\widetilde\varphi_{S,S_z}(\vec k) &=& 
   E\widetilde\varphi_{S,S_z}(\vec k) 
\,,\label{eq:69}\end{eqnarray}
for fixed and given $S$ and $S_z$ rather than carrying out 
the spin summations in Eq.(\ref{eq:50}) explicitly .
In this representation, the only spin-off-diagonal operator in
Eq.(\ref{eq:57}) is diagonal, \textit{i.e.}
\begin{eqnarray}
   \vec\sigma_1\vec\sigma_2&=& 2S(S+1) - 3 =
   \left\{\begin{array}{rll}
     +1, & \mbox{ for } S=1, &\mbox{ triplet}, \\
     -3, & \mbox{ for } S=0, &\mbox{ singlet}.\end{array} \right. 
\end{eqnarray}
It depends on the taste, whether one works first in configuration
space and subsequently Fourier transforms to momentum space,
or whether one solves directly Eq.(\ref{eq:69}) in momentum space. 
At the end one has $\widetilde\varphi_{S,S_z}(\vec k)$,
which can be unitarily transformed back to the light-cone
wave function according to
\begin{eqnarray} 
\lefteqn{\hspace{-2em}
   \psi_{h_1 h_2}(x,\vec k_{\!\perp})= 
   \frac{\sqrt{A(k_z(x),\vec k_{\!\perp})}}{\sqrt{x(1-x)}} 
}\nonumber\\ &\times&
   \sum_{s_1,s_2} 
   \langle h_1 h_2|\Omega^\dagger(k_z(x),\vec k_{\!\perp})|s_1 s_2\rangle 
\\ &\times&   
   \sum_{S,S_z}\textstyle
   \langle S,S_z\vert \frac12 s_1 \frac12\ s_2 \rangle 
   \widetilde\varphi_{S,S_z}(k_z(x),\vec k_{\!\perp})
\,,\nonumber\end{eqnarray}   
see  also Eqs.(\ref{eq:7}) and (\ref{eq:21}).
Note that the light cone wave function is a superposition 
of singlet and triplet.

\section{Perspectives}
I have promised a technical paper.
Sometimes technicalities are important.

Most of the above work relates to back ground knowledge on atomic 
and Dirac theory and the results speak for themselves.  

Instead of discussing the results to some detail,
I will highlight only some important aspects:
\begin{itemize}
\item[$\bullet$]
   Most remarkably, the fine and hyperfine interaction depends 
   on the coupling constant
   only through the potential energy $V(r)$.
\item[$\bullet$]
   One faces a seemingly non relativistic potential energy 
   which is directly related to the light cone 
   wave functions with all their wonderful advantages to calculate 
   cross sections, structure functions, distribution amplitudes,
   and the like.
\item[$\bullet$]
   In fact, one has the wonderful advantage of applying 
   phenomenological approaches working with potential energies,
   and transforming their solutions to light cone wave functions.
   One can then calculate dynamic quantities like cross sections.
   Thus far, this was possible only with a lot of hand waving.
\item[$\bullet$]
   This opens a broad avenue of model tailoring and
   comparison to experiment.
\item[$\bullet$]
   It seems one has gotten the cookie and the cake.
\end{itemize}

\begin{appendix}
\section{Fourier transform of the kinetic energy}
\label{asec:ke}
The kinetic energy is defined in Eq.(\ref{ieq:23})
\begin{eqnarray*} 
   T(\vec k;\vec k') &=& \phantom{-}
   \frac{k ^2}{2m_r}\delta^{(3)}(\vec k-\vec k') 
\,.\end{eqnarray*}
The Fourier transform according to Eq.(\ref{eq:47}) is
\begin{eqnarray} && 
   \int\!\!d^3\!\vec r' T(\vec r;\vec r')\Psi(\vec r') =
\\ &&=  
   \int\!\!d^3\!\vec r' \!\int\!\!d^3\!\vec k \!\int\!\!d^3\!\vec k' 
   \frac{\mathrm{e}^{i(\vec k\vec r-\vec k'\vec r')}}{(2\pi)^3}
   T(\vec k;\vec k') \Psi(\vec r')
\nonumber\\ &&= 
   \int\!\!d^3\!\vec r' \!\int\!\!d^3\!\vec k  
   \frac{\vec k^2}{2m_r}
   \frac{\mathrm{e}^{i\vec k(\vec r-\vec r')}}{(2\pi)^3}
   \Psi(\vec r') =
   \frac{-{\vec\nabla}^2}{2m_r} \Psi(\vec r) 
\,.\nonumber\end{eqnarray} 

\section{Fourier transform of the potential energy}
\label{asec:pe}
Inspection of Eq.(\ref{ieq:43}) and Eq.(\ref{ieq:45}) reveals
the potential energy $U(\vec k;\vec k')$ to depend on the momenta
through the combinations $\vec q=\vec k-\vec k'$, $(\vec k\vec k')$ and
$(\vec k\wedge\vec k')$. It is therefore convenient to introduce
sum and difference, \textit{i.e.}  
\begin{eqnarray} 
   \begin{array} {rcl c rcl} 
     \vec k\phantom{'} &=& \vec p + \frac{\vec q}{2} , &\qquad&
     \vec q &=& \phantom{\frac12(}\vec k - \vec k' ,
   \\ 
     \vec k'&=& \vec p - \frac{\vec q}{2} , &&
     \vec p &=& \frac12(\vec k+\vec k')   .
   \end{array} 
\end{eqnarray} 
The typical combinations are then
\begin{eqnarray} \textstyle
   \begin{array} {rcl} 
   \vec k\cdot\vec k' &=& \textstyle \vec p ^2 - \frac{\vec q ^2}{4}
\,,\\ \textstyle 
   \vec k\wedge\vec k' &=& \textstyle \vec q\wedge\vec p
\,,\\ \textstyle 
   \vec k \vec r - \vec k'\vec r' &=& \textstyle 
   \vec p (\vec r-\vec r')  + \frac{\vec q}{2}(\vec r+\vec r')
\,.\end{array} 
\end{eqnarray} 
The general Fourier transform Eq.(\ref{eq:47}) is then replaced by
\begin{eqnarray} 
   H (\vec r;\vec r') &=&
   \int\! 
   \textstyle{\frac{d^3\!\vec p\ d^3\!\vec q}{(2\pi)^3}}\ %
   \mathrm{e}^{i\vec p(\vec r-\vec r')}
   \mathrm{e}^{i\frac{\vec q}{2}(\vec r+\vec r')}
   H (\vec q;\vec p)
\,,\end{eqnarray} 
\textit{i.e.} the kernel must be expressed 
in terms of $\vec q$ and $\vec p$.

\subsection{The central potential}
\label{asec:c}
According to Eq.(\ref{ieq:43}) and (\ref{ieq:45}) the central potential is
\begin{eqnarray*} 
   U_\mathrm{c}(\vec q;\vec p)\equiv U(q)= -
   \frac{\alpha_c(q)}{q^2}\frac{R(q)}{2\pi^2} 
\,.\end{eqnarray*}
Define the function $V(r)$, 
\begin{eqnarray} 
   V(r)  =  \int\!\!d^3\!\vec q\ \mathrm{e}^{i\vec q\vec r} U(q)
\,,\end{eqnarray}
and get  
\begin{eqnarray} 
   V_\mathrm{c}(\vec r;\vec r') &=&  
   \int\!\!d^3\!\vec q\ \mathrm{e}^{\frac i2 \vec q(\vec r+\vec r')}
   U(q)
   \!\int\!\!d^3\!\vec p 
   \frac{\mathrm{e}^{i\vec p(\vec r-\vec r')}}{(2\pi)^3}
\,,\nonumber\\ &=& 
   \delta^{(3)}(\vec r-\vec r') V(r)
\,.\label{aeq:67}\end{eqnarray} 
Folding with the wave function,
\begin{eqnarray*} 
   V_\mathrm{c}\otimes\Psi\equiv\int\!\!d^3 \vec r'
   V_\mathrm{c}(\vec r;\vec r') \ \Psi(\vec r') =
   V(r)\Psi(\vec r)  = V\Psi 
\,,\end{eqnarray*} 
produces the potential energy $V(r)$ as a multiplicative factor.

\subsection{General rules}
\label{asec:r}
Consider first the simpler case of $U_a(\vec q;\vec p)\equiv\vec q\,U(q)$.
According to Eq.(\ref{aeq:67}) its Fourier transform is
\begin{eqnarray*} 
   V_\mathrm{a}(\vec r;\vec r') =
   \int\!\!d^3\!\vec q \!\int\!\!d^3\!\vec p 
   \ \vec qU(q)
   \frac{\mathrm{e}^{i(\vec p (\vec r-\vec r')+
   \frac12\vec q(\vec r+\vec r'))}}{(2\pi)^3}
\,.\end{eqnarray*} 
Integrating first over $\vec p$ gives a Dirac delta function
\begin{eqnarray*} 
   V_\mathrm{a}(\vec r;\vec r') =
   \delta^{(3)}(\vec r-\vec r')
   \int\!\!d^3\!\vec q\ \vec q U(q)\ \mathrm{e}^{i\vec q\vec r} 
\,.\end{eqnarray*} 
The $\vec q$ can be substituted by the partial 
$\vec q \Longrightarrow -i\vec \nabla_r,$
which gives 
$V_\mathrm{a}(\vec r;\vec r') =
   \delta^{(3)}(\vec r-\vec r')\ (-i\vec \nabla_r V(r))
.$ 
Folding gives
\begin{eqnarray*} 
   V_\mathrm{a}\otimes\Psi = -i\vec \nabla_V(V\Psi) 
\,,\end{eqnarray*} 
where $\nabla_V$ acts only on $V(r)$.
It is clear how to generalize this to arbitrary powers
$U_a(\vec q;\vec p)= (\vec q)^n\, U(q)$.

Next, consider $U_b(\vec q;\vec p)\equiv\vec p\, U(q)$. 
Take the Fourier transform first over $\vec q$, thus
\begin{eqnarray*} 
   V_\mathrm{b}(\vec r;\vec r') &=&
   \int\!\!d^3\!\vec p \ \vec p 
   \frac{\mathrm{e}^{i(\vec p (\vec r-\vec r')}}{(2\pi)^3}
   \int\!\!d^3\!\vec q \ U(q) 
   \mathrm{e}^{\frac12\vec q(\vec r+\vec r'))} 
\\ &=&
   V\left(\textstyle\frac12(\vec r+\vec r')\right)
   \int\!\!d^3\!\vec p \ \vec p\ %
   \frac{\mathrm{e}^{i(\vec p (\vec r-\vec r')}}{(2\pi)^3}
\,.\end{eqnarray*} 
The $\vec p$ can be substituted by the partial 
$\vec p \Longrightarrow -i\vec \nabla_r,$
which gives 
$V_\mathrm{b}(\vec r;\vec r') = V(r)\ %
   (-i\vec \nabla_r \delta^{(3)}(\vec r-\vec r'))
.$ 
In the folding, the derivative of the delta function generates
a derivative of the wave function $\Psi(\vec r)$,
\begin{eqnarray*} 
   V_\mathrm{b}\otimes\Psi = -i\vec \nabla_\Psi(V\Psi) 
\,,\end{eqnarray*} 
where $\nabla_\Psi$ acts only on $\Psi(\vec r)$.
It is clear how to generalize this to arbitrary powers
$U_b(\vec q;\vec p)= (\vec p)^n\, U(q)$.

These findings can be cast into the \textbf{general rules}:
\begin{enumerate}
\item
Consider the kernel of a general integral equation:
\begin{eqnarray}
    U_g(\vec k;\vec k') = f(\vec k,\vec k')\,U(|\vec k-\vec k'|)
\end{eqnarray} 
\item
Introduce $\vec p$ and $\vec q$ according to 
\begin{eqnarray} 
   \begin{array} {rcl c rcl} 
     \vec k\phantom{'} &=& \vec p + \frac{\vec q}{2} , &\qquad&
     \vec q &=& \phantom{\frac12(}\vec k - \vec k' ,
   \\ 
     \vec k'&=& \vec p - \frac{\vec q}{2} , &&
     \vec p &=& \frac12(\vec k+\vec k')   .
   \end{array} 
\end{eqnarray}
\item
Express  $U_g$ in terms of $\vec p$ and $\vec q$:
\begin{eqnarray}
    U_g(\vec q;\vec p) = F(\vec q,\vec p)\,U(q)
\end{eqnarray} 
\item
Substitute $\vec q$ and $\vec p$ according to:
\begin{eqnarray}
   \begin{array} {rcl c rcl} 
    \vec q &\Longrightarrow& -i\vec \nabla_V ,   \\
    \vec p &\Longrightarrow& -i\vec \nabla_\Psi .  
   \end{array} 
\end{eqnarray} 
\item 
Get the folded Fourier transform by
\begin{eqnarray}
    V_\mathrm{g}\otimes\Psi = F(-i\vec \nabla_V,-i\vec \nabla_\Psi)(V\Psi)
\,.\end{eqnarray} 
\end{enumerate}
This suffices to Fourier transform all functions of interest.

\section{The different terms of the interaction}
\label{asec:DiffTe}
According to Eqs.(\ref{ieq:43},\ref{ieq:45})  
the hyperfine term is 
\begin{eqnarray} \textstyle 
   U_\mathrm{hf}(\vec k;\vec k') &=& \textstyle 
   (\vec k-\vec k')^2 U(q)\ %
   \frac{(-\vec\sigma_1\vec\sigma_2 )}{6m_1 m_2}
\,,\\ \textstyle
   U_\mathrm{hf}(\vec q;\vec p) &=& \textstyle 
   q ^2 U(q)\ %
   \frac{(-\vec\sigma_1\vec\sigma_2 )}{6m_1 m_2}
\,,\nonumber\\ \textstyle
   V_\mathrm{hf}\otimes\Psi &=& \textstyle -
   \nabla_V^2 \left(V\Psi\right)
   \frac{(-\vec\sigma_1\vec\sigma_2 )}{6m_1 m_2}
.\nonumber\end{eqnarray}
The kinetic term is
\begin{eqnarray} \textstyle 
   U_\mathrm{K}(\vec k;\vec k') &=& \textstyle 
   \frac{(\vec k+\vec k ')^2}{4m_1 m_2}\,U(q)
\,,\\ \textstyle
   U_\mathrm{K}(\vec q;\vec p) &=& \textstyle 
   p ^2 \frac{U(q)}{m_1m_2}   
\,,\nonumber\\ \textstyle
   V_\mathrm{K}\otimes\Psi &=& \textstyle -
   \nabla_\psi^2 \frac{\left(V\Psi\right)}{m_1m_2}   
.\nonumber\end{eqnarray}
The Darwin term is
\begin{eqnarray} \textstyle 
   U_\mathrm{D}(\vec k;\vec k') &=& \textstyle 
   \frac{\vec k\cdot\vec k '}{4m_{rs}^2} U(q)
\,,\\ \textstyle
   U_\mathrm{D}(\vec q;\vec p) &=& \textstyle 
   \left(q^2-\frac14 p ^2\right) \frac{U(q)}{4m_{rs}^2} 
\,,\nonumber\\ \textstyle
   V_\mathrm{D}\otimes\Psi &=& \textstyle -
   \left(\nabla_V^2-\frac{1}{4}\nabla_\psi^2\right)
   \frac{\left(V\Psi\right)}{4m_{rs}^2} 
.\nonumber\end{eqnarray}
The typical spin orbit term is
\begin{eqnarray} \textstyle 
   U_\mathrm{so_1}(\vec k;\vec k') &=& \textstyle 
   \frac{i\vec\sigma_1\cdot\vec k\wedge\vec k '} {4m_1} U(q)\ %
   \Big[\frac{1}{m_r }+\frac{1}{m_1}\Big]
\,,\label{aeq:70}\\ \textstyle
   U_\mathrm{so_1}(\vec q;\vec p) &=& \textstyle 
   \frac{i\vec\sigma_1\cdot\vec q\wedge\vec p} {4m_1} U(q)\ %
   \Big[\frac{1}{m_r }+\frac{1}{m_1}\Big]
\,,\nonumber\\ \textstyle
   V_\mathrm{so_1}\otimes\Psi &=& \textstyle - 
   \frac{i\vec\sigma_1\cdot\vec \nabla_V\wedge\vec \nabla_\Psi} {4m_1}\ %
   \left(V\Psi\right)\ %
   \Big[\frac{1}{m_r }+\frac{1}{m_1}\Big]
.\nonumber\end{eqnarray}
The $L^2$-term is 
\begin{eqnarray} \textstyle 
   U_\mathrm{L}(\vec k;\vec k') &=& \textstyle -
   \frac{-\left(\vec k\wedge\vec k'\right)^2}{(4 m_1 m_2)^2}U(q) 
\,,\label{aeq:71}\\ \textstyle
   U_\mathrm{L}(\vec q;\vec p) &=& \textstyle -
   \frac{\left(\vec q\wedge\vec p\right)^2}{(4 m_1 m_2)^2} U(q) 
\,,\nonumber\\ \textstyle
   V_\mathrm{L}\otimes\Psi &=& \textstyle - 
   \frac{\left(\vec \nabla_V\wedge\vec \nabla_\Psi\right)^2}{(4 m_1 m_2)^2}\ %
   \left(V\Psi\right) 
\,.\nonumber\end{eqnarray}  
The hyperfine-$L^2$ term is 
\begin{eqnarray} \textstyle 
   U_\mathrm{hfL}(\vec k;\vec k') &=& \textstyle -
   \frac{\left(\vec k\wedge\vec k'\right)^2}{4 m_1 m_2} U(q)\, 
   \frac{\vec\sigma_1\vec\sigma_2}{6 m_1 m_2}
\,,\label{aeq:72}\\ \textstyle
   U_\mathrm{hfL}(\vec q;\vec p) &=& \textstyle -
   \frac{\left(\vec q\wedge\vec p\right)^2}{4 m_1 m_2} U(q)\, 
   \frac{\vec\sigma_1\vec\sigma_2}{6 m_1 m_2}
\,,\nonumber\\ \textstyle
   V_\mathrm{hfL}\otimes\Psi &=& \textstyle \phantom{-} 
   \frac{\left(\vec \nabla_V\wedge\vec \nabla_\Psi\right)^2}{4 m_1 m_2}\ %
   \left(V\Psi\right)\ %
   \frac{\vec\sigma_1\vec\sigma_2}{6 m_1 m_2}
\,.\nonumber\end{eqnarray}  
The spin-orbit Darwin term is 
\begin{eqnarray} \textstyle 
   U_\mathrm{soD}(\vec k;\vec k') &=& \textstyle  
   \frac{i\left(
   \vec\sigma_1+\vec\sigma_2\right)\cdot\vec k\wedge\vec k '}{4 m_1 m_2}
   \left(\vec k\cdot\vec k '\right)\, U(q)\ %
   \frac{1}{4 m_1 m_2} 
\,,\label{aeq:73}\\ \textstyle
   U_\mathrm{soD}(\vec q;\vec p) &=& \textstyle  
   \frac{i\left(
   \vec\sigma_1+\vec\sigma_2\right)\cdot\vec q\wedge\vec p}{4 m_1 m_2}
   \left(q ^2-\frac{1}{4}p ^2\right)\, U(q)\ %
   \frac{1}{4 m_1 m_2} 
\,,\nonumber\\ \textstyle
   V_\mathrm{soD}\otimes\Psi &=& \textstyle   
   \frac{i\left(
   \vec\sigma_1+\vec\sigma_2\right)\cdot
   \vec \nabla_V\wedge\vec \nabla_\Psi}{4 m_1 m_2}
   \left(\nabla_V ^2-\frac{1}{4}\nabla_\Psi ^2\right)\, 
   \frac{\left(V\Psi\right)}{4 m_1 m_2} 
,\nonumber\end{eqnarray}
finally, which completes taking the Fourier transforms.

The potential $V(r)$ is strictly spherically symmetric. With
\begin{eqnarray}   
   \vec \nabla V(r) = \frac{\vec r}{r}\frac{d V}{dr}
\,,\end{eqnarray}  
and with the usual orbital angular momentum
\begin{eqnarray}  
   \vec L = -i \vec r\wedge\vec \nabla
\,,\end{eqnarray}  
one can simplify the above equations considerably. With
\begin{eqnarray*}   
   \left(\vec \nabla_V\wedge\vec \nabla_\Psi\right) \left(V\Psi\right) = 
   \left(\vec \nabla V\wedge\vec \nabla \Psi\right) = 
   \Big[\frac{1}{r}\frac{dV}{dr}\Big] 
   \Big[\vec r\wedge\vec \nabla \Psi\Big] 
\,,\end{eqnarray*} 
one can substitute 
in Eqs.(\ref{aeq:70})--(\ref{aeq:73})
\begin{eqnarray*}   
   \left(\vec \nabla_V\wedge\vec \nabla_\Psi\right) \Longrightarrow i
   \Big[\frac{1}{r}\frac{d}{dr}\Big]_V \vec L_\Psi 
\,,\end{eqnarray*}  
and replace them by 
\begin{eqnarray}   
   V_\mathrm{so_1}\otimes\Psi &=& \textstyle  
   \Big[\frac{1}{m_r }+\frac{1}{m_1}\Big]
   \frac{\vec\sigma_1\cdot\vec L_\Psi} {4m_1}\ %
   \Big[\frac{1}{r}\frac{d}{dr}\Big]_V 
   V\Psi 
,\\ \textstyle
   V_\mathrm{L}\otimes\Psi &=& \textstyle 
   \frac{\left(\vec L_\Psi\right)^2}{(4 m_1 m_2)^2}\ %
   \Big[\frac{1}{r^2}\frac{d^2}{dr^2}\Big]_V 
   V\Psi 
,\\ \textstyle
   V_\mathrm{hfL}\otimes\Psi &=& \textstyle 
   \frac{\vec\sigma_1\vec\sigma_2}{6 m_1 m_2}
   \frac{\left(\vec L_\Psi\right)^2}{4 m_1 m_2}\ %
   \Big[\frac{1}{r^2}\frac{d^2}{dr^2}\Big]_V 
   V\Psi 
,\\ \textstyle
   V_\mathrm{soD}\otimes\Psi &=& \textstyle  
   \frac{\left(
   \vec\sigma_1+\vec\sigma_2\right)\cdot\vec L_\Psi}{\left(4 m_1 m_2\right)^2}
   \Big[\frac{1}{r}\frac{d}{dr}\Big]_V 
   \left(\frac{1}{4}\nabla_\Psi ^2-\nabla_V ^2\right) 
   V\Psi  
,\end{eqnarray} 
a very suggestive form indeed.
 
\section{The non-local central potential}
\label{asec:6}
One can define the kernel of a spherically symmetric potential 
in which the Fourier approximation has \textbf{not} been made.
Restricting to the `1' in Eq.(\ref{eq:28}), 
one gets from Eq.(\ref{eq:23}): 
\begin{eqnarray} 
&&\lefteqn{\hspace{-3ex}  
   \widetilde U(\vec k,\vec k')= - 
   \frac{\alpha_c(Q)}{2\pi^2 Q^2} \frac{R(Q)}{4m_1m_2m_r}  
}\\ &\times&
   \sqrt{\frac{E_1(k)E_2(k)}{E_1(k)+E_2(k)}\ \ (E_1(k )+m_1)(E_2(k )+m_2)} 
\nonumber\\ &\times&
   \sqrt{\frac{E_1(k')E_2(k')}{E_1(k')+E_2(k')}(E_1(k')+m_1)(E_2(k')+m_2)} 
\,.\nonumber\end{eqnarray}
It is plainly impossible to find an analytical Fourier transform 
of this, except when applying series expansions 
in $\vec k\,^2/m^2$, \textit{i.e.} 
\begin{eqnarray} 
   E(k) &=& m\sqrt{1+\frac{\vec k\,^2}{m^2}} 
\nonumber\\  &\simeq& 
   m\left[1 + \frac12\left(\frac{\vec k\,^2}{m^2}\right) -
   \frac18\left(\frac{\vec k\,^2}{m^2}\right)^2 + \dots\right]
\,,\end{eqnarray}
but then all the beauty of the present approach gets lost.
In the Fourier approximation, Eq.(\ref{eq:26}),
only the first term is included.
\end{appendix}

\end{document}